\documentclass[12pt]{article}
\usepackage{latexsym}
\usepackage{amsmath}
\usepackage{epsfig}
\title{How Large is the ``Natural'' Magnetic Moment?}
\author{Barry R. Holstein\\
Department of Physics-LGRT\\
University of Massachusetts\\
Amherst, MA  01003}
\begin{document}
\begin{titlepage}
\maketitle
\begin{abstract}
The ``natural'' magnetic moment of a particle of spin $S$ is
generally assumed to be that given by the Belinfante conjecture
and has the value $g=1/S$ for its gyromagnetic ratio.  Thus, for
the spin 1/2 electron we find the Dirac value $g_{\rm e}=2$.
However, in the standard model the charged W boson, a spin one
particle, is found to have the value $g_{W^+}=2$. We show how this
result comes about and argue that the ``natural'' value for any
particle of spin S should be $g=2$, independent of spin.
\end{abstract}
\end{titlepage}

\section{Introduction}

One of the great successes of Dirac theory is the prediction of
the magnetic moment of the electron.  Usually this is written in
terms of the gyromagnetic ratio or g-factor, which is defined via
the relation
\begin{equation}
\vec{\mu}={eg\over 2m}\vec{S}\label{def}
\end{equation}
where $\vec{\mu}$ represents the magnetic moment of a particle of
mass $m$, charge $e$, and spin $\vec{S}$.  Then for spin 1/2 the
simple Dirac theory makes the prediction $g=2$, which is the value
found experimentally (except for very small corrections from
photon loop effects) for the electron, as well as for its standard
model partners $\mu$ and $\tau$\cite{dgh}.  Feynman argued that
the value $g=2$ could be generated in an intuitive fashion by
generalizing the Schr\"{o}dinger equation for a spin 0 system of
mass $m$ and charge $e$ interacting with an external vector
potential $A^\mu=(\phi,\vec{A})$---
\begin{equation}
i{\partial\over \partial
t}\psi=\left({(-i\vec{\nabla}-e\vec{A})^2\over
2m}+e\phi\right)\psi\label{eq:eq:sp}
\end{equation}
to the form
\begin{equation}
i{\partial\over \partial t}\psi=\left({1\over 2m
}\vec{\sigma}\cdot(-i\vec{\nabla}-e\vec{A})\vec{\sigma}\cdot(-i\vec{\nabla}-e\vec{A})
+e\phi\right)\psi\label{eq:so}
\end{equation}
in the case of spin 1/2\cite{fey}.  Using the Pauli matrix
identity
\begin{equation}
\sigma_i\sigma_j=\delta_{ij}+i\epsilon_{ijk}\sigma_k
\end{equation}
the spin 1/2 Schr\"{o}dinger equation can be written in the
alternative form
\begin{equation}
i{\partial\over \partial
t}\psi=\left({(-i\vec{\nabla}-e\vec{A})^2\over 2m}-{e\over 2m
}\vec{\sigma}\cdot\vec{\nabla}\times\vec{A}+e\phi\right)\psi
\end{equation}
wherein the Hamiltonian is that of the simple spinless system
accompanied by a magnetic moment interaction with g=2.

Of course, for charged spin 1/2 systems other than the
$e,\mu,\tau$, {\it e.g.}, the proton or neutron, there exist large
deviations from the Dirac value---$g_{\rm p}=5.58,\,\, g_{\rm
n}=-3.82$ on account of loop effects associated with the strong
interactions\cite{bdh}.  However, these are not fundamental
systems, but rather are bound states of constituent quarks.

It is then an interesting question to ask whether there is a
corresponding ``natural'' value for the g-factor of fundamental
systems having spin other than 1/2.  One answer to this question
was given long ago by Belinfante, who evaluated the magnetic
moment of a simple spin 3/2 system and observed that the g-factor
was 2/3.  Combining this with known results for Dirac (spin 1/2)
and Proca (spin 1) systems for which the g-factors are 2 and 1
respectively, Belinfante proposed that for a system of spin $S$,
$g_S=1/S$\cite{bel}, and this has become known as the Belinfante
conjecture.  This result was proven for arbitrary spin by Case and
others nearly five decades ago\cite{cas} and more recently by
Hagen and Hurley\cite{hh} and is based on the assumption that the
interaction of the spinning system with the electromagnetic field
is generated by the simple ``minimal substitution''\cite{min}---
\begin{equation}
i\nabla_\mu\longrightarrow \pi_\mu\equiv
i\nabla_\mu-eA_\mu\label{eq:ms}
\end{equation}
that is known from classical electrodynamics to generate the
interactions of a charged particle with an external vector
potential\cite{jj}.  However, in order to check Belinfante's
proposal, we are limited by the fact that the only other
manifestations of charged particles which do not interact strongly
are the $W^\pm$ bosons, which have unit spin and therefore would
be expected to have $g_W^{\rm Belinfante}=1$.  In the tree level
standard model, however, the charged W-boson is found to have
$g_W^{\rm exp}=2$, due to the requirement that it is also a gauge
boson for the electroweak interaction\cite{gau}. Below we shall
show how this feature comes about and will argue that in fact the
``natural'' value for the g-factor of a fundamental system of spin
$S$ is $g_S=2$---{\it independent of $S$!}.  This is not a new
suggestion, and rather has been reached by a number of authors in
recent years.  In particular the work of Ferrara, Porrati, and
Telegdi shows that Compton scattering from targets of mass $m$ and
arbitrary spin violates unitarity at photon energy $\omega_i\sim
m$ unless $g_S=2$\cite{fpt}, while for those readers looking for a
broad and very interesting historical summary as well as some of
the constraints posed by general relativity, the article by
Pfister and King is required reading\cite{pk}.  We shall give
below a more limited set of arguments, which, however, we believe
are more than enough to buttress the case.

In the next section then we review how the g-factor can be
identified in a given relativistic Lagrangian, using the cases of
spin 1/2 and spin 1 as examples.  (An alternative approach, based
on a nonrelativistic reduction, is presented in the Appendix.)  In
section 3 we demonstrate why the charged W-boson has a g-factor
which differs from that suggested by Belinfante and cite specific
arguments why one might expect the prediction $g=2$ to be
generally valid for systems of arbitrary spin.  Finally, we
summarize our results in a brief concluding section.

\section{ The Belinfante Conjecture}

In order to motivate the Belinfante conjecture, we need to see how
to extract the g-factor from a given relativistic Lagrangian. This
can be done in a number of ways.  A standard method is to use a
nonrelativistic reduction, as demonstrated in the Appendix.
However, one can also identify the g-factor directly by isolating
the magnetic interaction in the Lagrangian, as we show here. We
begin by reviewing how the Dirac value---$g=2$---arises for spin
1/2.
\subsection{S=1/2}
The well-known Dirac Lagrangian for a free spin 1/2 particle is
\begin{equation}
{\cal L}=\bar{\psi}(x)(i\not\!{\nabla}-m)\psi(x)
\end{equation}
Making the minimal substitution---Eq. \ref{eq:ms}---this becomes
for the case of a charged system
\begin{equation}
{\cal L}=\bar{\psi}(x)(i\not\!{\nabla}-e\not\!\!{A}-m)\psi(x)
\end{equation}
and we can identify the interaction Lagrangian by picking out the
piece of the Lagrangian proportional to $A^\mu$
\begin{equation}
{\cal L}_{\rm
int}=-eA^\mu(x)\bar{\psi}(x)\gamma_\mu\psi(x)\label{eq:in}
\end{equation}
By use of the free Dirac equation for the fields
\begin{equation}
\psi(x)={i\over m}\not\!{\nabla}\psi(x)
\end{equation}
Eq. \ref{eq:in} can be rewritten as
\begin{equation}
{\cal L}_{\rm int}={ie\over
2m}A^\mu\left[\bar{\psi}\gamma_\mu\gamma_\nu\nabla^\nu\psi-
(\nabla^\nu\bar{\psi})\gamma_\nu\gamma_\mu\psi\right]
\end{equation}
Then using the matrix identity
\begin{equation}
\gamma_\mu\gamma_\nu=\eta_{\mu\nu}-i\sigma_{\mu\nu}
\end{equation}
the interaction Lagrangian may be written in the so-called Gordon
form as\cite{min}
\begin{equation}
{\cal L}_{\rm int}={e\over
2m}A^\mu(x)\bar{\psi}(x)i\overleftrightarrow{\nabla}_\mu\psi(x)+{e\over
2m}A^\mu(x)\nabla^\nu(\bar{\psi}(x)\sigma_{\mu\nu}\psi(x))\label{eq:co}
\end{equation}
where we have defined
$$ D(x)\overleftrightarrow{\nabla}_\mu F(x)
\equiv D(x)\nabla_\mu F(x)-(\nabla_\mu D(x)) F(x)$$ The first
component of Eq. \ref{eq:co} is a convective term which is not of
interest to our present focus.  Rather we examine the second term,
which involves the total derivative, and observe that it can be
rewritten in the form
\begin{equation}
{\cal L}_{\rm int}=-{e\over
4m}F^{\mu\nu}\psi(x)\sigma_{\mu\nu}\psi(x)
\end{equation}
where we have integrated by parts and used the fact that
$\sigma_{\mu\nu}$ is antisymmetric in the indices $\mu\nu$.
Finally, noting that $F^{ij}=-\epsilon_{ij\ell}B^\ell$ and
$\sigma_{ij}=2\epsilon_{ijk}S_k$, where
$$\vec{S}=\left(\begin{array}{cc}
{1\over 2}\vec{\sigma}& 0\\
0&{1\over 2}\vec{\sigma}\end{array}\right)$$ is the spin operator,
we isolate the magnetic interaction---
\begin{equation}
{\cal L}_{\rm mag}={e\over
m}\bar{\psi}(x)\vec{S}\cdot\vec{B}\psi(x)
\end{equation}
and read off the well-known result $g=2$.

Of course, this prediction is not expected to be exact.  In the
case of the electron, photon loops make small ${\cal O}(\alpha)$
corrections, while, in the case of the proton, strong interaction
corrections yield large modifications of ${\cal O}(1)$.  Such
corrections, usually called the ``anomalous'' magnetic
moment---$\kappa$---can be accounted for phenomenologically by
inclusion of a so-called Pauli term in the Dirac
equation\cite{pau}---
\begin{equation}
(i\not\!{\nabla}-e\not\!\!{A}-{e\kappa\over
2m}\sigma_{\mu\nu}F^{\mu\nu}-m)\psi(x)=0
\end{equation}

As mentioned above, it has been proven rigorously that, for
arbitrary spin, inclusion of the electromagnetic interaction by
the minimal substitution yields the ``natural'' value for the
magnetic moment given by the Belinfante conjecture\cite{cas}.  The
proofs are rather formal, and we eschew the temptation to
reproduce them here.  Rather we shall examine only the case of
unit spin, in order to demonstrate how higher spins are handled.
This is an important case, however, because there exists a
fundamental charged particle that is analogous to the electron in
that it does not participate in the strong interactions and
therefore may be expected to carry its ``natural'' value of the
magnetic moment. This is the charged W-boson, which is the carrier
of the weak interaction, and can be used to check of these ideas.
Before discussing the $W^\pm$, however, we demonstrate how the
minimal prediction for a spin one magnetic moment comes about.

\subsection{Spin 1}

The Lagrangian which describes a free neutral spin one system is
that given by Proca\cite{pro}
\begin{equation}
{\cal L}={1\over 2
}B^{\alpha}(x)[\eta_{\alpha\beta}(\Box+m^2)-\nabla_{\beta}\nabla_{\alpha}]
B^\beta(x)
\end{equation}
and in the case of a spin 1 particle which is charged, we can
introduce the electromagnetic interaction as before by making the
minimal substitution---
\begin{equation}
{\cal
L}=B^{\alpha\dagger}(x)[\eta_{\alpha\beta}((\nabla+ieA)^\mu(\nabla+ieA)_\mu+m^2)-
(\nabla+ieA)_\beta(\nabla+ieA)_\alpha))]B^\beta(x)
\end{equation}
We isolate the single photon piece of the interaction
Lagrangian---
\begin{equation}
{\cal L}_{\rm int}=ieA^\mu(x)
B^{\alpha\dagger}(x)[\eta_{\alpha\beta}\overleftrightarrow{\nabla}_\mu
-\eta_{\beta\mu}\nabla_\alpha]B^\beta(x)-\eta_{\alpha\mu}(\nabla_\beta
B^{\alpha\dagger}(x))B^\beta(x)
\end{equation}
which, as before, can be rewritten as a linear combination of
total derivative and $\overleftrightarrow{\nabla}$ pieces as
\begin{eqnarray}
{\cal L}_{\rm int}&=&ieA^\mu
B^{\alpha\dagger}(x)[\eta_{\alpha\beta}\overleftrightarrow{\nabla}_\mu+{1\over
2}\eta_{\mu\beta}\overleftrightarrow{\nabla}_\alpha+{1\over
2}\eta_{\mu\alpha}\overleftrightarrow{\nabla}_\beta]B^\beta(x)\nonumber\\
&+&{ie\over
2}(\eta_{\mu\beta}\nabla_\alpha-\eta_{\mu\beta}\nabla_\alpha)(B^{\alpha\dagger}(x)B^\beta(x))
\end{eqnarray}
Neglecting the convective component, we focus on the total
derivative term, which, integrating by parts, assumes the form
\begin{equation}
{\cal L}_{\rm int}(x)=-{ie\over
2}F^{\mu\nu}(x)(B_\nu^\dagger(x)B_\mu(x)-B_\mu^\dagger(x)B_\nu(x))
\end{equation}
Defining matrix elements of the spin operator via\cite{sp}
\begin{equation}
B^\dagger_i B_j-B^\dagger_j
B_i=-i\epsilon_{ijk}<f|S_k|i>\label{eq:mm}
\end{equation}
the interaction Lagrangian becomes
\begin{equation}
{\cal L}_{\rm int}=e\vec{B}\cdot <f|\vec{S}|i>\label{eq:be}
\end{equation}
Finally, dividing by the factor 2m, which accounts for the
normalization of the unit spin states, we find the Belinfante
result for the g-factor---$g_{S=1}=1$.

\subsection{Charged W-Boson}

As discussed above, the electron magnetic moment agrees with its
``natural'' value given by the Belinfante conjecture up to small
terms due to photon loop corrections, but how about the charged
W-boson, which is the unit-spin analog of the electron in that
there can be no strong interaction corrections?  From Eq.
\ref{eq:be}, we would expect a g-factor having the value unity,
but in the tree level standard model the correct number is
predicted to be twice this value, and this prediction is confirmed
experimentally---$g_{W^\pm}=2.20\pm 0.20$\cite{pdg}. What is going
on, and why is such a large shift to be expected?

The answer can be found in the simple Lagrangian which describes
the charged W and the requirement that it be a Yang-Mills
field---{\it i.e.}, that the electroweak interaction is a gauge
theory\cite{gau}.  This means that the spin one Lagrangian which
contains the charged-W has the Proca form---
\begin{equation}
{\cal L}=-{1\over 4}(\vec{U}_{\mu\nu})^2+{m^2\over 2}\vec{U}_\mu^2
\end{equation}
but the SU(2) field tensor $\vec{U}_{\mu\nu}$ contains an
additional term on account of the required gauge
invariance\cite{gau}
\begin{equation}
\vec{U}_{\mu\nu}=\pi_\mu\vec{U}_\nu-\pi_\nu\vec{U}_\mu-ig\vec{U}_\mu\times\vec{U}_\nu
\end{equation}
where $g$ is the SU(2) electroweak coupling constant.  This
``extra'' term in the field tensor is responsible for the
interactions involving three and four W-bosons.  For our purpose,
however, we pick out only the term involving a pair of charged
W-bosons
\begin{eqnarray}
{\cal L}_{WW}&=&-{1\over
2}W^{+\dagger}_{\mu\nu}W^{+\mu\nu}-{1\over
2}W^{-\dagger}_{\mu\nu}W^{-\mu\nu}+{m^2\over 2}(W^{+\dagger}_\mu
W^{+\mu}+W^{-\dagger}_\mu
W^{-\mu})\nonumber\\
&-&gW^{0\mu\nu}(W^{+\dagger}_\mu W^+_\nu -W^{-\dagger}_\mu
W^-_\nu)\label{ww}
\end{eqnarray}
where here the field tensors involving the W-boson are of the
simple form
$$W_{\mu\nu}=\pi_\mu W_\nu-\pi_\nu W_\mu$$
We see that there appears to be an additional triple-W coupling.
However, this is illusory.  In the standard model there is no
neutral W-boson.  Rather the field $W^0$ is a linear combination
of the photon and neutral $Z$-boson fields,
\begin{equation}
W^0_\mu=\cos\theta_WZ^0_\mu+\sin\theta_W A_\mu
\end{equation}
with the Weinberg angle $\theta_W$ determining the
mixing\cite{wki}.  Since the combination $g\sin\theta_W$ is equal
to the electric charge $e$, the WW Lagrangian in Eq. \ref{ww}
assumes the form
\begin{eqnarray}
{\cal L}_{WW}&=&-{1\over
2}W^{+\dagger}_{\mu\nu}W^{+\mu\nu}-{1\over
2}W^{-\dagger}_{\mu\nu}W^{-\mu\nu}+{m^2\over 2}(W^{+\dagger}_\mu
W^{+\mu}+W^{-\dagger}_\mu
W^{-\mu})\nonumber\\
&-&eF_{\mu\nu}(W^{+\dagger\mu}W^{+\nu}-W^{-\dagger\mu}W^{-\nu})+\ldots\label{eq:sm}
\end{eqnarray}
where the terms indicated by the ellipses involve couplings to the
$Z$-boson.  By use of the identity Eq. \ref{eq:mm}, we can rewrite
the last piece of Eq. \ref{eq:sm} as
\begin{equation}
\delta{\cal L}_{WW}^{(2),\rm
int}=e(<f|\vec{S}|i>_{W^+}-<f|\vec{S}|i>_{W-})\cdot \vec{B}
\end{equation}
which is seen to have the form of an anomalous magnetic moment
and, when added to the previously discussed contribution to the
magnetic moment from the first line of Eq. \ref{eq:sm}, increases
the predicted g-factor from its Belinfante value---$g_{W^\pm}^{\rm
Belinfante}=1$---to its standard model value---$g_{W^\pm}^{\rm
sm}=2$.

\section{What is the ``Natural'' g-factor?}

In the previous section we have noted two very different results
concerning a ``natural'' value of the g-factor.  In the first, the
simple minimal substitution was shown to agree with the Belinfante
conjecture---$g=1/S$---while in the second an "extra" term
required by gauge invariance gave the result $g=2$ for the spin 1
$W^\pm$ gauge boson and agrees with the proposal that the
"natural" value should $g_S=2$, independent of spin.  These
predictions coincide in the case of spin 1/2, but differ for
higher spin.  So far, we have dealt only with the case of spin 1/2
and spin one, and we need additional input in order to deal with
higher spin.  It would be nice to be able to use experiment to
decide the case, but this is not possible for there exist only two
cases of charged particles which do not participate in the strong
interactions--the electron (or its partners the muon or tau) and
the W-boson.  Thus extension of our ideas beyond spin one requires
theoretical input. Below we discuss in turn at least three such
theoretical reasons which suggest that the correct answer is
$g_S=2$.

\subsection{Compton Scattering at High Energy}

The first argument comes from study of Compton scattering from a
spin-$S$ target at high energy\cite{wei}, \cite{fpt}.  First
consider the case of spin one.  The simple Proca interaction
yields the Feynman rules for photon interactions
\begin{eqnarray}
PP\gamma_1:
&=&-ie(p_f+p_i)_\mu g_{\alpha\beta}+ieg_{\beta\mu}p_{f\alpha}+ieg_{\alpha\mu}p_{i\beta}\nonumber\\
PP\gamma\gamma:
&=&ie^2(2g_{\mu\nu}g_{\alpha\beta}-g_{\mu\alpha}g_{\nu\beta}-g_{\mu\beta}g_{\nu\alpha})
\end{eqnarray}
and, for generality, we append an anomalous moment contribution of
the form
\begin{equation}
PP\gamma_2:
=-ie(g-1)[g_{\alpha\mu}(p_{f\beta}-p_{i\beta})+g_{\beta\mu}(p_{i\alpha}-p_{f\alpha})]
\end{equation}

\begin{figure}
\begin{center}
\epsfig{file=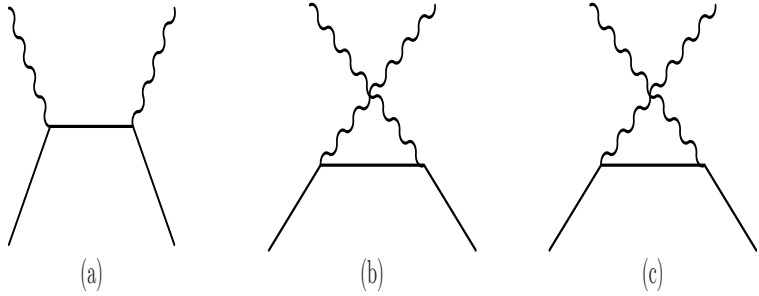,height=4cm,width=10cm} \caption{Diagrams
relevant to Compton scattering. }
\end{center}
\end{figure}

One can now perform the (somewhat tedious) evaluation of the three
lowest order Compton scattering diagrams shown in Figure 1,
yielding the result
\begin{eqnarray}
{\rm Amp}&=&e^2\{2\epsilon_A\cdot\epsilon_B\left[{\epsilon_1\cdot
p_1\epsilon_2\cdot p_2\over p_1\cdot k_1}-{\epsilon_1\cdot
p_2\epsilon_2\cdot p_1\over p_1\cdot
k_2}-\epsilon_1\cdot\epsilon_2\right]\nonumber\\
&-&g\left[\epsilon_A\cdot
[\epsilon_2,k_2]\cdot\epsilon_B\left({\epsilon_1\cdot p_1\over
p_1\cdot k_1} -{\epsilon_1\cdot p_2\over p_1\cdot k_2}\right)
-\epsilon_A\cdot[\epsilon_1,k_1]\cdot\epsilon_B\left({\epsilon_2\cdot
p_2\over p_1\cdot k_1}
-{\epsilon_2\cdot p_1\over p_1\cdot k_2}\right)\right]\nonumber\\
&-&g^2\left[{1\over 2p_1\cdot
k_1}\epsilon_A\cdot[\epsilon_1,k_1]\cdot[\epsilon_2,k_2]\cdot\epsilon_B
-{1\over 2p_1\cdot
k_2}\epsilon_A\cdot[\epsilon_2,k_2]\cdot[\epsilon_1,k_1]\epsilon_B\right]\nonumber\\
&-&{(g-2)^2\over m^2}\left[{1\over 2p_1\cdot
k_1}\epsilon_A\cdot[\epsilon_1,k_1]\cdot
p_1\epsilon_B\cdot[\epsilon_2,k_2]\cdot p_2\right.\nonumber\\
&-&\left.{1\over 2p_1\cdot
k_2}\epsilon_A\cdot[\epsilon_2,k_2]\cdot p_1
\epsilon_B\cdot[\epsilon_1,k_1]\cdot p_1\right]\}
\end{eqnarray}
where we have defined
$$S\cdot[Q,R]\cdot T\equiv S\cdot QT\cdot R-S\cdot RT\cdot Q.$$
The intriguing terms here are those on the last two lines, which
are proportional to the factor $1/m^2$.  They arise from the Born
diagrams via the $k_\alpha k_\beta$ piece of the spin-one
propagator
\begin{equation}
D_{\alpha\beta}(k)={i\over k^2-m^2}(-g_{\alpha\beta}+{k_\alpha
k_\beta\over m^2})
\end{equation}
and reveal that if we take the limit as the charge stays fixed and
the mass becomes small the Compton amplitude will diverge,
violating unitarity at a photon energy $\omega_i\sim m$ {\it
unless the gyromagnetic ratio has the value $g=2$!}  Remarkably,
the condition $g=2$ has been demonstrated by Ferrara, Porrati, and
Telegdi to assure the absence of $1/m^2$ terms {\it for arbitrary
spin}\cite{fpt}, which is certainly suggestive that the
``natural'' value of the g-factor should be $g_S=2$.

\subsection{The GDH Sum Rule}

The Gerasimov, Drell, Hearn sum rule relates the anomalous
magnetic moment of a system of spin $S$ to a weighted integral
over polarized photon cross sections.  It was originally derived
for the nucleon and has the form
\begin{equation}
{\alpha\kappa^2\over m^2}={1\over \pi^2}\int_0^\infty
{d\omega\over \omega}(\sigma_{3\over 2 }(\omega)-\sigma_{1\over 2
}(\omega))
\end{equation}
where here $\kappa$ is the anomalous magnetic moment, while
$\sigma_{3\over 2}(\omega)$, $\sigma_{1\over 2}(\omega)$ are the
laboratory frame scattering cross sections involving a polarized
nucleon target and circularly polarized photons whose helicity is
parallel, antiparallel to the target spin.  It is derived simply
by assuming analyticity of the Compton amplitude together with
crossing symmetry and convergence and is a very fundamental sum
rule about which much has been written\cite{dre}.  Recent
experimental studies have shown that it is satisfied in the case
of a nucleon target\cite{dre}.The GDH sum rule is also closely
connected to the celebrated Bjorken sum rule\cite{bjs}
\begin{equation}
\int_0^1 (g_1^{\rm p}(x)-g_1^{\rm n}(x))={1\over 6}g_A
\end{equation}
which relates the difference of polarized proton and neutron
inclusive electron scattering structure functions to the axial
decay constant of the neutron.  For these reasons the GDH sum rule
can be considered to be very fundamental and generally valid for
arbitrary targets even though it has been experimentally verified
only for the nucleon.  The generalization of the GDH sum rule to
the case of arbitrary spin $S$ has been given by
Weinberg\cite{wei}, who showed that it has the form
\begin{equation}
{\alpha\over 4M^2}(g_S-2)^2={1\over
\pi^2}\int_0^\infty{d\omega\over
\omega}[\sigma_{1+S}(\omega)-\sigma_{1-S}(\omega)]
\end{equation}
where here the cross sections are scattering cross sections
circularly polarized photons with their polarization parallel and
anti parallel to that of a maximally polarized target of spin $S$.
If as in the nucleon case, we define what is measured here as the
{\it anomalous magnetic moment}, then we see that this anomalous
moment is defined in terms of the difference of the experimental
g-factor from a bare value---$g_{\rm bare}=2$---again supporting
the suggestion that the "natural" value of the g-factor should be
$g_S=2$, independent of spin.

\subsection{Graviton Scattering and Factorization}

\begin{figure}
\begin{center}
\epsfig{file=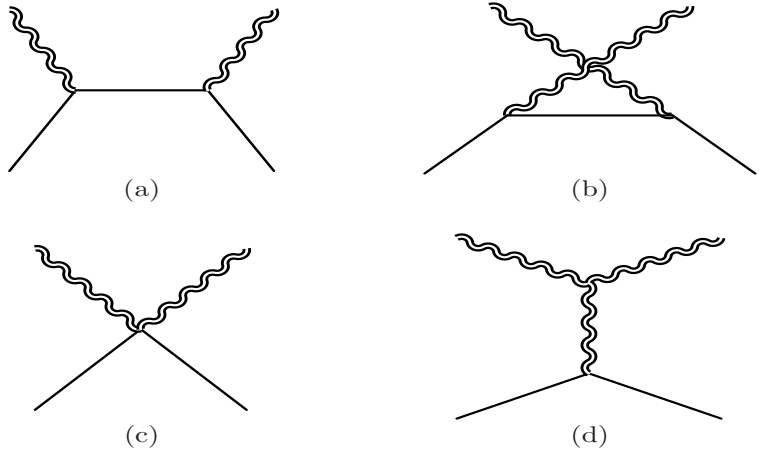,height=6cm,width=10cm} \caption{Diagrams
relevant for gravitational Compton scattering. }
\end{center}
\end{figure}

Using powerful (string-based) techniques, which simplify
conventional quantum field theory calculations, it has recently
been demonstrated that the elastic scattering of gravitons from a
target of arbitrary spin must factorize\cite{fac}, a feature that
had been noted ten years previously by Choi et al. based on gauge
theory arguments\cite{so}.  That is, a (in harmonic gauge)
graviton is a particle of spin 2 whose polarization vector
$\epsilon_{\mu\nu}$ can be written as a simple product of the
corresponding spin one photon polarization vectors---
$$\epsilon_{\mu\nu}^{\pm 2}=\epsilon_\mu^{\pm 1}\epsilon_\nu^{\pm 1}$$
The elastic scattering of gravitons from a target of arbitrary
spin is constructed from the four diagrams shown in Figure 2,
consisting of two Born diagrams, a seagull term, and the graviton
pole diagram.  The factorization theorem asserts that for
scattering from a target of spin $S$, the graviton scattering
amplitude can be written in the form
\begin{equation}
\epsilon_f^{*\alpha\beta}M^G_{\alpha\beta;\mu\nu}\epsilon_i^{\mu\nu}
=F\times(\epsilon_f^{*\alpha}A^C_{\alpha;\mu}(S)\epsilon_i^\mu)\times(\epsilon_f^{*\alpha}
A^C_{\alpha;\mu}(0)\epsilon_i^\mu)
\end{equation}
where $M^G_{\alpha\beta;\mu\nu}$ is the elastic graviton
scattering amplitude, $A^C_{\alpha\mu}(S)$ is the elastic Compton
amplitude from a target of spin $S$, and F is the kinematic factor
\begin{equation}
F={G\over 4\pi\alpha^2}{p_1\cdot k_1p_1\cdot k_2\over k_1\cdot
k_2}
\end{equation}
with $\alpha=e^2/4\pi$ and $G$ being the fine structure and Newton
constants respectively.  The full graviton scattering amplitude
could in principle have terms proportional to $1/m^2$ from the
Born diagrams and the propagator of the spin-$S$ system. However,
{\it this does not occur}. This can be seen in the case of a spin
one system from the form of the half-off-shell energy-momentum
tensor, which determines the gravitational coupling
\begin{eqnarray}
<p_2,\lambda|T_{\mu\nu}|p_1,\epsilon_A>&=&\epsilon_{A\lambda}(p_{2\mu}p_{1\nu}+p_{1\nu}p_{2\mu})
+g_{\mu\nu} p_{1\lambda}\epsilon_A\cdot p_2\nonumber\\
&-&p_{1\lambda}(p_{2\mu}\epsilon_{A\nu}+p_{2\nu}\epsilon_{A\mu})-\epsilon_A\cdot
p_2(p_{1\mu}g_{\lambda\nu}+p_{1\nu}g_{\lambda\mu})\nonumber\\
&+&(p_2\cdot
p_1-m^2)(g_{\lambda\mu}\epsilon_{A\nu}+\epsilon_{A\mu}g_{\lambda\nu}-g_{\mu\nu}\epsilon_{A\lambda})
\end{eqnarray}
Taking $p_1$ to be physical---$p_1^2=m^2$---but $p_2$ to be an
off-shell intermediate state momentum, we find that when
$T_{\mu\nu}$ is contracted with the $1/m^2$ piece of the spin one
propagator---${p_2^\alpha p_2^\lambda\over m^2}$---the result is
{\it independent} of $m^2$---
\begin{equation}
{p_2^\alpha p_2^\lambda\over m^2}
<p_2,\lambda|T_{\mu\nu}|p_1,\epsilon_A>=p_2^\alpha(g_{\mu\nu}\epsilon_A\cdot
p_2-\epsilon_{A\mu}p_{2\nu}-\epsilon_{A\nu}p_{2\mu})
\end{equation}
The coefficient $1/m^2$ in the spin one propagator has cancelled.
Thus no term proportional to $1/m^2$ survives in the Born
amplitude and this vanishing of terms which diverge as
$m\rightarrow 0$ can be shown to be a general property regardless
of the spin of the target.  According to the factorization
condition, the vanishing of $1/m^2$ terms in the gravitational
amplitude can only result from the vanishing of such terms in the
corresponding Compton amplitude, which we have already seen occurs
{\it only} if the value $g_S=2$ is chosen, so from an additional
standpoint we see that the ``natural'' value for the g-factor is
$g=2$.

\section{Conclusions}

Above we have examined the question of whether there exists a
``natural'' value for the gyromagnetic ratio of a fundamental
particle of spin $S$.  One might think that this question was
answered long ago when Belinfante conjectured and others proved
the assertion that when minimal substitution is used in order to
generate the electromagnetic interactions os a system of spin $S$,
the resulting g-factor is given by $g=1/S.$  However, while this
prediction is consistent with the case of the electron, we showed
that in the standard model of electroweak interactions the only
other example of a charged fundamental particle---the $W^\pm$
boson---does not agree with this prediction.  Rather, due to gauge
invariance, the result $g_W=2$  is found.  We then went on to
argue that, while there are no additional experimental cases, the
``natural'' value $g=2$ for all systems is supported by at least
three theoretical arguments:
\begin{itemize}
\item [a)] Use of $g_S=2$ guarantees the vanishing of terms
proportional to $1/m^2$ in the Compton scattering amplitude from a
system of spin $S$, whose presence would violate unitarity at the
very low energy scale $\omega\sim m$.
\item [b)] The fundamental GDH sum rule allows experimental
measurement of the {\it anomalous} magnetic moment for a system of
arbitrary spin $S$, where the anomalous moment is
$$\kappa=g_{\rm exp}-g_S$$ with $g_S=2$ being the "natural"
value of the g-factor.
\item [c)] Based on the result that the graviton
scattering amplitude from a system of spin-S must factorize into a
product of Compton amplitudes, the vanishing of $1/m^2$ terms in
the graviton amplitude is found to result from the choice $g_S=2$
in the corresponding Compton amplitude.
\end{itemize}
Although for simplicity we shall not discuss them here there exist
even more reasons for this assertion\cite{fpt},\cite{pk}.
\begin{itemize}
\item [d)] The only known example of a completely consistent
theory of interacting particles of spin greater than two is string
theory, and for open strings it is possible to obtain the exact
equations of motion for massive charged particles of arbitrary
spin, moving in a constant, external electromagnetic background.
This procedure yields $g_S=2$ for all spins\cite{str}.
\item [e)] The classical relativistic equation of motion of the polarization
vector $S_\mu$ in a homogeneous external electromagnetic field is
given by the Bargmann, Michel, Telegdi or BMT equation\cite{bmt}
\begin{equation}
{dS^\mu\over d\tau}={eg\over 2m}F^{\mu\nu}S_\nu+{e\over
2m}(g-2){dx^\mu\over d\tau} F^{\nu\lambda}S_\nu {dx_\lambda\over
d\tau}
\end{equation}
which simplifies for $g_S=2$ independent of the spin of the
particle.
\item [f)] In general relativity the $G\rightarrow 0$ limit of
the Kerr-Newman metric describing spacetime around a charged
spinning mass results in an electromagnetic field in flat space
with $g=2$.  Additional arguments from general relativity can be
found in \cite{pk}.
\end{itemize}

For these and other reasons then, it appears that the ``natural''
value for the g-factor is $g_S=2$ regardless of spin. Since there
are no experimental manifestations of this result outside of the
electron and $W^\pm$-boson systems, this discussion could be
argued to be somewhat metaphysical, but the totality of the
arguments given above together with the standard model value of
the $W^\pm$ moment would seem to constitute a rather compelling
case.

\begin{center}
{\bf Acknowledgements}
\end{center}

This work was supported in part by the National Science Foundation
under award PHY 02-44801.  Thanks to Prof. A. Faessler and the
theoretical physics group at the University of T\"{u}bingen, where
this work was completed, for hospitality.

\section{Appendix: Nonrelativisitic Reduction}

As an alternative derivation of Belinfante's result, it is useful
to understand how to construct an effective nonrelativistic
Schr\"{o}dinger Hamiltonian from a given relativistic
Lagrangian\cite{hol}.  In this way, for example, one finds the
familiar Dirac prediction that for a fundamental spin 1/2 system
we have $g=2$.
\subsection{Spin 1/2}
In order to see how this prediction comes about, we begin with the
Dirac equation in the absence of electromagnetism
\begin{equation}
(i\not\!{\nabla}-m)\psi(x)=0
\end{equation}
where $\not\!{V}$ implies contraction of the four-vector $V_\mu$
with the Dirac matrices---$\gamma_0V_0-\vec{\gamma}\cdot\vec{V}$.
Now assume that we can account for coupling to electromagnetism
via the minimal substitution
\begin{equation}
i\nabla_\mu\longrightarrow \pi_\mu\equiv i\nabla_\mu-eA_\mu
\end{equation}
so that the Dirac equation becomes
\begin{equation}
(\not\!{\pi}-m)\psi(x)=0
\end{equation}
Using the conventional representation for the Dirac
matrices\cite{min}
\begin{equation}
\gamma_0=\left(\begin{array}{cc}1&0\\0&1\end{array}\right),\quad
\vec{\gamma}=\left(\begin{array}{cc}0&\vec{\sigma}\\-\vec{\sigma}&0\end{array}\right)
\end{equation}
we can write the Dirac equation as a set of two coupled equations
relating the upper ($\psi_a$) and lower ($\psi_b$) components of
the wavefunction.  That is, for a positive energy solution
\begin{equation}
\psi(\vec{x},t)=\left(\begin{array}{c}\psi_a(\vec{x})\\
\psi_b(\vec{x})\end{array}\right)\exp(-i(m+W)t)
\end{equation}
we have
\begin{eqnarray}
(m+W-e\varphi)\psi_a-\vec{\sigma}\cdot\vec{\pi}\psi_b&=&m\psi_a\nonumber\\
\vec{\sigma}\cdot\vec{\pi}\psi_a-(m+W-e\varphi)\psi_b&=&m\psi_b
\end{eqnarray}
Solving the second of these for the lower component we find
\begin{equation}
\psi_b={1\over 2m+W-e\varphi}\vec{\sigma}\cdot\vec{\pi}\psi_a
\end{equation}
and substitution into the first yields
\begin{equation}
\vec{\sigma}\cdot\vec{\pi}{1\over
2m+W-e\varphi}\vec{\sigma}\cdot\vec{\pi}\psi_a=(W-e\varphi)\psi_a
\end{equation}
which in the nonrelativistic limit---$W<<m$---becomes
\begin{equation}
\left[{1\over
2m}\vec{\sigma}\cdot\vec{\pi}\vec{\sigma}\cdot\vec{\pi}+e\varphi\right]\psi_a=W\psi_a
\end{equation}
and produces the effective Schr\"{o}dinger Hamiltonian
\begin{equation}
H_{\rm eff}={1\over
2m}\vec{\sigma}\cdot\vec{\pi}\vec{\sigma}\cdot\vec{\pi}+e\varphi
\end{equation}
Using the Pauli matrix identity
\begin{equation}
\sigma_a\sigma_b=\delta_{ab}+i\epsilon_{abc}\sigma_c
\end{equation}
this becomes the expected result
\begin{equation}
H_{\rm eff}={(\vec{p}^2-e\vec{A})^2\over 2m}-{e\over
2m}\vec{\sigma}\cdot\vec{B}+e\phi
\end{equation}
and comparison with the definition Eq. \ref{def} and the usual
expression for the interaction energy of a magnetic dipole
\begin{equation}
U=-\vec{\mu}\cdot\vec{B}
\end{equation}
yields the prediction $g=2$.

\subsection{Spin Zero}

In order to see how to handle the case of unit spin, it is useful
to first review the method to obtain the effective Schr\"{o}dinger
equation for the case of spin zero.  We begin with the
Klein-Gordon equation
\begin{equation}
(\Box+m^2)\phi=0\label{kg}
\end{equation}
and include electromagnetism via the minimal substitution, so that
Eq. \ref{kg} assumes the form
\begin{equation}
(\pi^2-m^2)\phi=0
\end{equation}
A problem here is that this equation is second order in time. Thus
instead write it as a pair of coupled first order equations---
\begin{eqnarray}
m\chi_\mu&=&\pi_\mu\phi\nonumber\\
\pi^\mu\chi_\mu&=&m\phi
\end{eqnarray}
Note then that the vector components $\chi_i$ have no time
development and can therefore be considered as a constraint---
\begin{equation}
\chi_i={1\over m}\pi_i\phi
\end{equation}
We have then the two coupled equations
\begin{eqnarray}
i{\partial\over \partial
t}\chi_0&=&(m+e\varphi)\phi+{\vec{\pi}^2\over
m}\phi\nonumber\\
i{\partial\over \partial t}\phi&=&(m+e\varphi)\chi_0
\end{eqnarray}
 Now write the equation in terms of a two component "spinor"
\begin{equation}
\rho=\left(\begin{array}
{c}\rho_a\\ \rho_b\end{array}\right)={1\over 2}\left(\begin{array}{c}\phi+\chi_0\\
\phi-\chi_0\end{array}\right)
\end{equation}
We have then the equation
\begin{equation}
i{\partial\over \partial
t}\rho=\left[m\tau_3+e\varphi+{\vec{\pi}^2\over
2m}(\tau_3+i\tau_2)\right]\rho
\end{equation}
Projecting out the positive energy solution via
$$\rho(\vec{x},t)=\rho(\vec{x})\exp(-i(m+W)t)$$
the lower component of the spinor equation then has the form
\begin{equation}
(m+W-e\varphi)\rho_b=-m\rho_b-{\vec{\pi}^2\over 2m}(\rho_a+\rho_b)
\end{equation}
and can be solved in the nonrelativistic limit to yield
\begin{equation}
\rho_b\simeq -{\vec{\pi}^2\over 4m^2}\rho_a
\end{equation}
Substitution into the top component of the spinor equation then
yields
\begin{equation}
W\rho_a\simeq[e\varphi+{\vec{\pi}^2\over 2m}-{\vec{\pi}^4\over
8m^3}]\rho_a
\end{equation}
We have then the effective Schr\"{o}dinger Hamiltonian
\begin{equation}
H_{\rm eff}=e\varphi+{(\vec{p}-e\vec{A})^2\over
2m}-{(\vec{p}-e\vec{A})^4\over 8m^3}+\ldots
\end{equation}
as expected.  Similar methods can be used in order to to treat the
more challenging unit spin problem, as we now demonstrate.

\subsection {Spin One}
 In the case of spin 1 the free particle equation of motion
is given by the Proca equation\cite{pro},
\begin{equation}
\nabla^\mu
U_{\mu\nu}+m^2U_\nu=(\Box+m^2)U_\nu-\nabla_\nu\nabla\cdot
U=0\label{proca}
\end{equation}
where $U_{\mu\nu}$ is the field tensor
$$U_{\mu\nu}=\nabla_\mu U_\nu-\nabla_\nu U_\mu$$
Of course, a spin one field should only have three degrees of
freedom, while the four-vector $U_\mu$ has four.  However, from
Eq. \ref{proca}, we find, taking the divergence, that
\begin{equation}
m^2\nabla\cdot U=0
\end{equation}
which, for non-massless particles, yields the desired constraint
$\nabla\cdot U =0$.  In order to include interactions with the
electromagnetic field we make the minimal
substitution---$i\nabla_\mu\rightarrow \pi_\mu$ as before,
yielding the equation of motion
\begin{equation}
\pi^\mu(\pi_\mu U_\nu-\pi_\nu U_\mu)-m^2U_\nu=0
\end{equation}
and the constraint equation
\begin{equation}
{1\over 2}[\pi^\nu,\pi^\mu]U_{\mu\nu}=-i{e\over
2}F^{\mu\nu}U_{\mu\nu}=m^2\nabla\cdot U
\end{equation}
In order to reduce this equation to a nonrelativistic form we need
to rewrite it in terms of a pair of coupled first order
equations\cite{bly}
\begin{eqnarray}
\pi^\mu U_{\mu\nu}-m^2U_\nu=0\nonumber\\
 U_{\mu\nu}=\pi_\mu U_\nu-\pi_\nu U_\mu
\end{eqnarray}
We note then that the degrees of freedom $U_0$ and $U_{ij}$ do not
contribute to the time development and can be considered as
constraints
\begin{eqnarray}
U_{ij}=\pi_iU_j-\pi_jU_i\nonumber\\
U_0=-{1\over m^2}\vec{\pi}\cdot\vec{\phi}
\end{eqnarray}
where we have defined $U_{i0}=m\phi_i$.  We find then the
dynamical equations
\begin{eqnarray}
i{\partial\over \partial
t}\vec{\phi}&=&e\varphi\vec{\phi}+m\vec{U}-{1\over
m}\vec{\pi}\times(\vec{\pi}\times \vec{U})\nonumber\\
i{\partial\over \partial t}\vec{U}&=&e\varphi\vec{U}
+m\vec{\phi}+{1\over m}\vec{\pi}\vec{\pi}\cdot\vec{\phi}
\end{eqnarray}
Defining the spin vectors
\begin{equation}
S_x=\left(\begin{array}{ccc}0&0&0\\0&0&-i\\0&i&0\end{array}\right),\quad
S_y=\left(\begin{array}{ccc}0&0&i\\0&0&0\\-i&0&0\end{array}\right),\quad
S_z=\left(\begin{array}{ccc}0&-i&0\\i&0&0\\0&0&0\end{array}\right)
\end{equation}
and representing the vector quantities $\vec{\phi},\vec{U}$ in
terms of three-component column vectors $\phi,U$, these equations
become
\begin{eqnarray}
i{\partial\over \partial t}\phi&=&e\varphi\phi+mU+{1\over
m}(\vec{S}\cdot\vec{\pi})^2U\nonumber\\
i{\partial\over \partial t}U&=&e\varphi U +m\phi-{1\over
m}(\vec{S}\cdot\vec{\pi})^2\phi +{\vec{\pi}^2\over m}\phi-{e\over
m}(\vec{S}\cdot\vec{B})\phi
\end{eqnarray}
As in the Klein-Gordon case we can write this as a six-component
spinor equation by defining
\begin{equation}
\rho=\left(\begin{array}{c}\rho_a\\
\rho_b\end{array}\right)=
{1\over 2}\left(\begin{array}{c}\phi+U\\
\phi-U\end{array}\right)
\end{equation}
so that the equation reads
\begin{equation}
i{\partial\over \partial t}\rho=[e\varphi+\tau_3m-i\tau_2{1\over m
}(\vec{S}\cdot\vec{\pi})^2+(\tau_3+i\tau_2){1\over
2m}(\vec{\pi}^2+e\vec{S}\cdot\vec{B})]\rho
\end{equation}
Projecting out the positive energy solution via
\begin{equation}
\rho(\vec{x},t)=\rho(\vec{x})\exp(-i(m+W)t)
\end{equation}
we can solve for the lower component in the nonrelativistic limit,
yielding
\begin{equation}
\rho_b\simeq -{1\over 2m}({\vec{\pi}^2\over 2m}-{e\over
2m}(\vec{S}\cdot\vec{B})-{1\over
m}(\vec{S}\cdot\vec{\pi})^2)\rho_a
\end{equation}
and substitution into the top component yields the equation
\begin{equation}
W\rho_a=\left[e\varphi +{\vec{\pi}^2\over 2m}-{e\over
2m}(\vec{S}\cdot\vec{B})-{1\over 2m}({\vec{\pi}^2\over 2m}-{1\over
m}(\vec{S}\cdot\vec{\pi})^2)^2\right]\rho_a
\end{equation}
We thus find the effective spin one Hamiltonian to be
\begin{equation}
H_{\rm eff}={(\vec{p}-e\vec{A})^2\over 2m}-{e\over
2m}\vec{S}\cdot\vec{B}-{(\vec{p}-e\vec{A})^4\over 8m^3}+\ldots
\end{equation}
so that minimal substitution has yielded a value for the magnetic
moment---$g=1=1/s$---which agrees with the Belinfante conjecture.


\begin{thebibliography}{99}
\bibitem{pau} An interesting historical footnote, as noted by Drechsel,
is that when Stern
originally proposed measurement of a possible deviation of the
proton moment from the Dirac prediction, Pauli assured him that
this was a waste of time.  It is thus ironic that this anomalous
magnetic moment piece is called the Pauli term.
\bibitem{sp}  One can see this result by using the
representations of the rest frame polarization vectors of the spin
1 bosons quantized along the $z$-direction
$$\vec{\epsilon}_{\pm 1}=\sqrt{1\over 2}(\hat{x}\pm i\hat{y}),
\qquad \vec{\epsilon}_0=\hat{z}$$
and verifying, for example that
$$1=<1,1|S_z|1,1>=-i(\vec{\epsilon}_1^*\times\vec{\epsilon}_1)_z$$
Similarly the other matrix elements can be checked.
\end{thebibliography}
\end{document}